\documentclass[paper]{JHEP3} 

\title{Holomorphic Currents and Duality in $N=1$ Supersymmetric Theories}

\author{Andrei Johansen\\The St. Petersburg Nuclear Physics Institute\\Gatchina, 188350, Russian Federation\\
E-mail: \email{ajohansen@bluecrestna.com}}

\abstract{Twisted supersymmetric theories on a 
product of two Riemann surfaces possess non-local holomorphic currents
in a BRST cohomology.
The holomorphic currents act as vector fields on the chiral ring.
The OPE's of these currents are invariant under the renormalization
group flow up to BRST-exact terms.
In the context of electric-magnetic duality, the algebra generated
by the holomorphic currents in the
electric theory is isomorphic to the one on the magnetic side.
For the currents corresponding to global symmetries
this isomorphism follows from 't Hooft anomaly matching conditions.
The isomorphism between OPE's of the currents 
corresponding to non-linear transformations of fields of matter imposes 
non-trivial conditions on the duality map of chiral ring.
We consider in detail the $SU(N_c)$ SQCD with matter in fundamental and adjoint representations,
and find agreement with the duality map
proposed by Kutasov, Schwimmer and Seiberg.}

\preprint{}

\begin{document}

\section{Introduction}

Holomorphy is one of the most powerful tools in supersymmetric theories in four dimensions
(for a review, see e.g. \cite{isreview,terning}).
It strongly constraints the phase structure and the nonperturbative behavior of the 
supersymmetric quantum systems.
In particular, the current understanding of electric-magnetic duality 
is essentially based on the analysis of
moduli spaces of vacua, chiral rings and holomorphic deformations
of the superpotentials (see, e.g. \cite{terning}).

In this paper we consider aspects of supersymmetric theories that are not entirely controlled
by chiral rings and holomorphy in the chiral field configuration space.
More specifically, we consider twisted $N=1$ supersymmetric theories with a
conserved $R$-current on a K\"ahler manifold
$M=\Sigma_1 \times \Sigma_2$,
where $\Sigma_1$ ($\Sigma_2$) are two-dimensional Riemann surfaces.
In general the twisted theory contains fields with fractional spins.
We construct non-local holomorphic currents on $\Sigma_1$ ($\Sigma_2$)
in a BRST cohomology \cite{algebras}.
These currents correspond to
the conserved (generalized) currents of the untwisted theory.
However, in contrast to the ordinary currents
the coefficients in the operator product expansions (OPE) of 
the holomorphic currents are invariant under 
renormalization group flow and holomorphic
in the coordinates on $\Sigma_1$($\Sigma_2$) up to BRST-exact terms.
In particular, the central charge in the OPE of
the holomorphic spin-2 current (associated with the
stress tensor) with itself coincides with the {\it infrared}
value of the $a$ central charge, defined as the coefficient of 
the Euler density in the trace anomaly in the 
curved space.
The central charge $a$ is conjectured 
\cite{cardy,jackOsborn,cappelliFL,cappelliLV,bastianelli,anselmiFGJ,anselmiEFJ,forte,anselmi,cappelliAGM}
to obey the four-dimensional
analog of Zamolodchikov's $c$-theorem \cite{zamolodchikov}.
The holomorphic currents act as vector fields on the chiral ring.
This fact has interesting consequences for electric-magnetic duality.

In the context of electric-magnetic duality the operator 
algebra generated by the OPE's 
of holomorphic currents in the
electric theory has to be isomorphic to the one on the magnetic side. 
We show that for the currents corresponding to global symmetries of the theory 
this isomorphism follows from 't Hooft anomaly matching
conditions.
The isomorphism between OPE's of the currents 
corresponding to non-linear transformations of fields of matter imposes 
non-trivial conditions on the duality map of the chiral ring.
We apply this construction
to the $SU(N_c)$ supersymmetric QCD \cite{emseiberg} 
and the Kutasov-Schwimmer models \cite{kutasov,kutsch}.
For these models there is a very simple duality map for holomorphic currents 
corresponding to global symmetries.
In particular, the central charges of these currents 
coincide in the electric and magnetic descriptions of the theory.
In the Kutasov-Schwimmer models we also construct holomorphic currents 
corresponding to 
{\it generalized} baryon currents that are associated with non-linear transformations of quark superfields.
The action of these currents on the chiral ring agrees in 
a non-trivial way with the duality map for
the chiral operators proposed in \cite{kutasov, kutsch, emmap}.
This observation provides an additional check of the duality in the 
Kutasov-Schwimmer models.

The paper is organized as follows.
In Section 2 we review and generalize the twisting of 
$N=1$ supersymmetric theories with a conserved $R$-current 
on K\"ahler manifold.
In Section 3 we define non-local currents that are 
holomorphic in the BRST cohomology and discuss their OPE algebra.
In Section 4 we apply this construction
to the $SU(N_c)$ SQCD with matter in fundamental and adjoint representations.
We conclude by summarizing the results of the paper.
Some possible generalizations of these results are also discussed.

\section{Twisted $N=1$ supersymmetric theory}

In \cite{twist} it has been shown that a supersymmetric theory on a K\"ahler manifold $M$
can be twisted to preserve one of the supersymmetry generators.
Such a twisting is essentially
equivalent to coupling of the theory to an appropriate external vector
$U(1)$ field $V_{\mu}$, that 
enters the same $N=1$ supergravity multiplet with the veirbein field $e_{\mu}^a$.
This vector field effectively changes spins of the quantum fields.
The unbroken supersymmetry
generator becomes a BRST generator $Q$ of the twisted theory.
For consistency at the quantum level one has to ensure that the vector field $V_{\mu}$ 
is coupled to a non-anomalous conserved $R$-current.

In this paper we focus on the theories twisted on $M=\Sigma_1\times\Sigma_2$, where 
$\Sigma_{1, 2}$ are Riemann surfaces.
The metric on $M$ is chosen to be block diagonal $g=g_{\Sigma_1}\oplus g_{\Sigma_2}$.
We denote the complex coordinates on $\Sigma_1$
and $\Sigma_2$ as $z$ ($\bar{z}$) and $w$ ($\bar{w}$) respectively.
The holonomy group of $M$ is $U(1)\times U(1)$.
This allows us to consider a partial twisting
\cite{reduction}.
For example, we can choose $\Sigma_1=T^2$ a torus with a flat metric and keep the fields untwisted
along $\Sigma_1$ and twisted along $\Sigma_2$. 
However, to simplify notations we will consider the
fully twisted version of the theory.

Consider a supersymmetric gauge theory with a gauge group $G$ and
chiral superfields $\Phi_i$
in $r_i$ representations of the gauge group.
The gauge multiplet of the twisted theory contains a gauge field
$A_\mu = \{A_z, A_w, A_{\bar{z}}, A_{\bar{w}}\}$,
a scalar fermion $\bar{\lambda}$
and a fermionic $(\bar{1}, \bar{1})$-differential 
$\bar{\lambda}_{\bar{z}\bar{w}}$,
a fermionic $(1, 0)$-differential $\chi_z$ and
a fermionic $(0, 1)$-differential $\chi_w$, and an auxiliary scalar field $D$.

The $R$-charges of the chiral superfields in 
the untwisted theory may be fractional.
Therefore, the twisted theory may contain fields 
with fractional spins.
In order to determine the spin content of a twisted supermultiplet $\Phi$
it is instructive to notice that
the $R$-current is a linear combination of the canonical $R$-current
(corresponding to
phase rotations of the Grassmann coordinates $\theta$ of 
the untwisted theory) and
flavor currents. 
A chiral multiplet twisted with the canonical $R$-current 
\cite{algebras} 
contains a scalar bosonic field $\phi$ ($\bar{\phi}$), a scalar fermion $\bar{\psi}$, the fermionic fields
$\psi_{\bar{z}}$, $\psi_{\bar{w}}$ and $\bar{\psi}_{zw}$ which are $(\bar{1}, 0)$,
$(0,\bar{1})$ and $(1,1)$-differentials respectively, and
the auxiliary bosonic fields transforming
$N_{\bar{z}\bar{w}}$ and $\bar{N}_{zw}$ as
$(\bar{1},\bar{1})$ and $(1,1)$-differentials respectively.

An additional twist with $2j$ positive flavor charge 
corresponds to an additional twist of $\phi$ and $\psi_{\bar{z}(\bar{w})}$
with $K^j$ and $\bar{\phi}$ and $\bar{\psi}$, $\bar{\psi_{zw}}$ with 
$K^{-j}$, 
where $K$ stands for the canonical bundle over $\Sigma_2$.
This follows from the fact that by
twisting a field with one unit of $R$-charge its spin changes by $1/2$.
We will assume that $K^j$ and $K^{-j}$ are well defined line bundles
over $\Sigma_2$ and hence, $j\chi$ is an integer, where $\chi$ 
stands for the Euler characteristic of $\Sigma_2$.
As a result, under the twisting the fields become differentials of 
fractional degrees as follows
\begin{equation} 
\phi \rightarrow (j, j),
~\psi_{\bar{w}} \rightarrow (j,\overline{1-j}),
~\psi_{\bar{z}} \rightarrow (\overline{1-j},j),
~N_{\bar{z}\bar{w}} \rightarrow (\overline{1-j}, \overline{1-j}),
\label{diffs}
\end{equation}
$$\bar{\phi} \rightarrow (\bar{j}, \bar{j}),
~\bar{\psi} \rightarrow (\bar{j},\bar{j}),
~\bar{\psi}_{zw} \rightarrow (1-j, 1-j), 
~\bar{N}_{zw} \rightarrow (1-j, 1-j).$$
Here $(j_1, j_2)$ stands for a differential of the degrees $j_1$ and
$j_2$ on the $\Sigma_1$ and $\Sigma_2$ respectively.
In these notations a differential $\omega$ on a surface $\Sigma$ 
of a degree $j$ ($\bar{j}$ ) transforms as 
$\omega(z, \bar{z}) dz^j\to\omega(z', \bar{z})dz'^j$
($\omega(z, \bar{z}) d\bar{z}^j\to\omega(z', \bar{z})d\bar{z'}^j$)
under $z\to z', \bar{z}\to\bar{z'}$.

In order to simplify notations, 
we do not show explicitly the degrees of the differentials.
Instead, we will keep notations corresponding to 
the canonical twisting so that
the additional twist is implemented by a non-trivial external $U(1)$ field. 

The twisted Lagrangian in the theory without any superpotential 
reads as follows \cite{twist}: 
$$L = {1 \over e^2}\sqrt{g} 
\{Q, {\rm Tr}[-i\bar{\lambda}g^{z\bar{z}}F_{z\bar{z}}
-i\bar{\lambda}g^{w\bar{w}}F_{w\bar{w}}+{1 \over 2}D\bar{\lambda}-{i\over 2}
\bar{\lambda}_{\bar{z}\bar{w}} F^{\bar{z}\bar{w}}]\}~+$$
\begin{equation}
\sum_i\sqrt{g} \{Q, -{1\over 2}\bar{\psi}^{\bar{z}\bar{w},i} 
N^i_{\bar{z}\bar{w}}+ ~
\bar{\phi}^i(D^{\bar{z}}\psi^i_{\bar{z}}+D^{\bar{w}}\psi^i_{\bar{w}})
-~i\bar{\phi}^i\bar{\lambda}\phi^i\},
\label{lag}
\end{equation}
where we left color indices implicit and $F$ stands for the strength
tensor of the gauge field, $e^2$ is the gauge coupling.
Here $Q$ is the scalar nilpotent generator of BRST
transformations which for the gauge multiplet
read 
\begin{equation}
QA_z = \chi_z, ~QA_w = \chi_w, ~
QA_{\bar{z}}=QA_{\bar{w}} =0,~
Q\chi_z=Q\chi_w = 0,
\label{brstgauge}
\end{equation}
$$ Q\bar{\lambda}=-D,
~Q\bar{\lambda}_{\bar{z}\bar{w}}=2iF_{\bar{z}\bar{w}},~QD =0$$
while for the chiral multiplet $\Phi^i$ we have
\begin{equation}
Q\phi^i = 0, ~Q\psi^i_{\bar{w}}=D_{\bar{w}}\phi^i,
~Q\psi^i_{\bar{z}}=D_{\bar{z}}\phi^i,~
QN_{zw}^i=D_{\bar{z}}\psi^i_{\bar{w}}-D_{\bar{w}}\psi^i_{\bar{z}}+
{1 \over 2}\bar{\lambda}_{\bar{z}\bar{w}}\phi^i,
\label{brstmatter}
\end{equation}
$$Q\bar{\phi}^i=\bar{\psi}^i,~ Q\bar{\psi}^i =0, ~
Q\bar{\psi}_{zw}^i= -2\bar{N}_{zw}^i,~ Q\bar{N}_{zw}^i=0,$$
where $D_{z(w)}$ and $D_{\bar{z}(\bar{w})}$ stand for covariant derivatives.

The ghost numbers can be taken equal to the $R$-charges of the fields.
Note the expression (\ref{lag}) is $Q$-exact.
Therefore, one can
perform computations of OPE's of $Q$-closed
operators in the weak coupling limit $e^2\to 0$.

In the case of a non-vanishing
superpotential $W \neq 0$ the contribution to the Lagrangian of the 
superpotential terms reads as
\begin{equation}
L_{sp}=\sqrt{g}[\sum_{i,j}
\psi_{\bar{z}}^i\psi_{\bar{w}}^j \partial_i\partial_jW(\phi)+
{1 \over 2}\sum_i N_{\bar{z}\bar{w}}^i\partial_i W(\phi)+
\{Q, \sum_i \bar{\psi}_{zw}^i\partial_i W(\bar{\phi})\}].
\label{superpotential}
\end{equation}
Here the gauge indices are implicit, and 
we assume that the superpotential $W$ does not violate the $R$-symmetry.
Therefore, the expression in the brackets transforms as a scalar.
One can check that $\{Q, L_{sp}\} = 0$.
Note that the last term in eq.
(\ref{superpotential}) is $Q$-exact and can be neglected in computations
in the $Q$-cohomology.
The chiral ring of the untwisted theory maps to a ring of local observables that are invariant under $Q$.
If $H^{2, 0}(M) \neq 0$ then it is also possible to construct non-local $Q$-closed operators 
as integrals over holomorphic 2-cycles in $M$ \cite{algebras}.
However, these observables are irrelevant for the discussion below.

\section{Holomorphic currents}

In \cite{algebras} it was shown that in the twisted theory one can define
BRST-closed currents on $\Sigma_1$ that are holomorphic 
up to $Q$-exact terms.
These currents generate chiral algebras\footnote{Chiral "projected algebras" 
have been also discussed in \cite{losev}
in more general context of higher-dimensional analogues of the $bc$ system of $2D$ RCFT.}
on $\Sigma_1$.
Here we consider a generalization of this construction.
Let us define a spin 1 current on $\Sigma_1$
(we omit the dependence on external metric)
\begin{equation}
j_z(g)=2\pi\sum_i\int_{\Sigma_2}
(-D_z\bar{\phi}^i g^i+\sum_j \bar{\psi}^i_{zw}
(\partial_j g^i)\psi^j_{\bar{w}}).
\label{J}
\end{equation}
Here $g^i = g^i(\phi)$ are arbitrary functions of $\phi^k$ 
transforming as
$R_i/2$-differentials on $\Sigma_2$ (also assume that $g^i$ do not 
depend on coordinates explicitly).
We also assume that $g^i$ transform according to the representation $r_i$ of
the gauge group.
By imposing the condition of the Q-closeness of $j_z(g)$ and using
the equations of motion one can easily get
\begin{equation}
\int_{\Sigma_2}\sum_i\sum_j\partial_j(g^i\partial_i W)\psi^j_{\bar{w}}= 0.
\label{totalQclose}
\end{equation}
Therefore, this condition is satisfied for
\begin{equation}
\sum_i g^i\partial_i W=0.
\label{Qclose}
\end{equation}
This condition means that the superpotential $W$ is invariant under
transformations $\phi^i\to\phi^i+g^i(\phi)$.

In general, the condition $Qj_z(g) = 0$ may be also violated by anomalies.
The condition of
absence of anomalies can be obtained at the one-loop level
\cite{algebras} and reads
\begin{equation}
\sum_i(\partial_ig^i)T(r_i)=0,
\label{anomaly}
\end{equation}
where $T(r_i)$ stands for the Dynkin index of 
the $r_i$ representation of the gauge group.
An important property of the current $j_z(g)$ is that
it is holomorphic in the Q-cohomology, i.e.
\begin{equation}
\partial_{\bar{z}}j_z(g) = \{Q,...\},
\label{holom}
\end{equation}
where $\{Q,...\}$ stands for Q-exact terms.
This fact can be checked by using the equations of motion.
eq. (\ref{holom}) may also suffer from an anomaly.
However, this anomaly exactly cancels for (\ref{anomaly}).

The coefficients in the OPE of two currents $j_z(g)$ and $j_z(g')$ 
are invariant under
the renormalization group modulo Q-exact terms \cite{algebras}.
Notice that the Lagrangian of the theory is a sum of 
various Q-exact terms and the Q-closed chiral contribution
from the superpotential. 
The superpotential is not renormalizable (see, e.g. \cite{terning}).
Therefore, the dependence on a renormalization group scale 
may appear only from $Q$-exact terms in the Lagrangian 
that may contribute only $Q$-exact terms to the OPE.

Let us consider the OPE of $j_z(g)(z, \bar{z})$
with an element of the chiral ring given by a polynomial $P(\phi)$.
We have 
\begin{equation}
j_z(g)(z, \bar{z})~P(u,\bar{u}, v, \bar{v})=
\label{jchiralOPE}
\end{equation}
$$-2\pi\int_{\Sigma_2} d^2 w \sum_i g^i(z, \bar{z}, w, \bar{w})
\partial_k P(u,\bar{u}, v, \bar{v})
\langle D_z\bar{\phi}^i(z, \bar{z}, w, \bar{w})
\phi^k(u,\bar{u}, v, \bar{v})\rangle+...=$$
$$-2\pi\int_{\Sigma_2} d^2w\sum_ig^i\partial_iP~
{1\over 4\pi^2}\partial_z{1 \over |z-u|^2+|w-v|^2}+...=
-{1\over 2}\sum_ig^i~\partial_iP{1\over z-u}+...$$
Here $z,\bar{z}$ and $u,\bar{u}$ stand for
coordinates on $\Sigma_1$, and $w,\bar{w}$ and $v,\bar{v}$ 
are coordinates on $\Sigma_2$.
In eq. (\ref{jchiralOPE}) we omitted $Q$-exact and regular terms.
One can see that the current $j_z(g)$ acts as a vector field
in the chiral ring generated by polynomials of chiral fields $\Phi^i$.

As shown in \cite{algebras} the holomorphic currents $j^a$ corresponding to a flavor group $G_F$ 
generate the Kac-Moody algebra (in this case the functions $g^i$ are linear in the fields of matter)
\begin{equation}
j^a_z(z,\bar{z})j^b_z(0,0)=
{k\delta^{ab}\over z^2}+{if^{abc}\over z}j^c(0,0)+
{\rm regular~and~}Q{\rm -exact~terms}
\label{kacmoody}
\end{equation}
where the central charge $k\propto (1-h)$ is constant and 
$h$ is the genus of $\Sigma_2$.

In general, for non-linear functions $g^i$ the algebra generated by the OPE
is more complicated.
Here we focus only on the most singular "central" terms.
In the limit of vanishing gauge coupling one has
\begin{equation}
j_z(g)(z,\bar{z})~j_z(g')(u,\bar{u})=
(2\pi)^2\int_{\Sigma_2}d^2w\int_{\Sigma_2}d^2v
\sum_{i,j}~(\partial_jg^i)(z,w)~(\partial_ig'^j)(u,v)
\label{genKacmoody}
\end{equation}
$$(\langle\phi^{j}(z, w)\partial_u\bar{\phi^i}(u, v)\rangle
\langle\phi^{i}(u,v)\partial_z\bar{\phi^j}(z,w)\rangle-$$
$$\langle\psi_{\bar{w}}^{j}(z,w)\bar{\psi}_{uv}^i(u,v)\rangle
\langle\psi_{\bar{v}}^{i}(u,v)\bar{\psi}_{zw}^j(z,w)\rangle)+
...$$
The propagators read as follows
$$\langle\phi^j(z, w)\bar{\phi^i}(u, v)\rangle = 
{\delta^{ij}\over\Delta_{\Sigma_1}^{({R_i\over 2})} +
\Delta_{\Sigma_2}^{({R_i\over 2})}}\delta^2(z-u)\delta^2(w-v),$$
$$\langle\psi_{\bar{w}}^j(z,w)\psi_{\bar{v}}^i(u,v)\rangle =
{\delta^{ij}\over\Delta_{\Sigma_1}^{({R_i\over 2})} +
\Delta_{\Sigma_2}^{(1-{R_i\over 2})}}\delta^2(z-u)\delta^2(w-v),$$
where $\Delta_{\Sigma}^{(r)}$ stands for 
the Laplace operator on $\Sigma$ for differentials of degree $r$.
The dependence of the chiral fields $\phi^i$ on antiholomorphic coordinates 
$\bar{w}$ and $\bar{v}$ can be dropped
since $D_{\bar{w}}\phi^i$ is $Q$-exact.
However, in the presence of external metric the dependence on holomorphic coordinates 
$w$ and $v$ is important.

By expanding $g^i$ in powers of $w-v$ one gets contributions of local operators of
different dimensions.
For the contribution of the operator of the lowest dimension one can drop the dependence
of $g$ and $g'$ on the holomorphic coordinates.
In this case the contributions of the non-zero modes of the Laplace operators 
on $\Sigma_2$ to eq. (\ref{genKacmoody})
are canceled due to the underlying supersymmetry.
The contributions of zero modes give
\begin{equation}
-{1\over (z-u)^2}\sum_{i,j} {\rm index}(R_i/2)
~\partial_jg^i~\partial_ig'^j=
-{\chi \over 2(z-u)^2}\sum_{i,j} 
(\partial_j g^i)(\partial_i g'^j)(1-R_i),
\label{zeroModeCentral}
\end{equation}
where ${\rm index}(R_i/2)$ stands for the index of the Dirac operator acting on
a $r$-differential on $\Sigma_2$, and 
$\chi=2(1-h)$ is the Euler characteristic of $\Sigma_2$.
In eq. (\ref{zeroModeCentral}) we used the Riemann-Roch formula 
\cite{dirac} to compute ${\rm index}(r)$
$${\rm index}(r)=h^{0,r}-h^{0,1-r}=(1-2r)(1-h),$$
where $h$ is a genus of $\Sigma_2$.
The "central" operator in eq.(\ref{zeroModeCentral}) is an element of the chiral ring.
In particular, for linear functions $g^i=q_i\phi^i$ and 
$g'^i={q'}_i\phi^i$ the currents $j_z(g)$ and $j_z(g')$ correspond
to $U(1)_q$ and $U(1)_{q'}$ flavor groups, and
$q(\phi^i)=q_i$ and $q'(\phi^i)=q'_i$.
In this case one gets the central term
$$-{\chi \over 2(z-u)^2}\sum_i q_iq'_i(1-R_i)$$
that is proportional to the 't Hooft anomaly $\langle qq'R\rangle$.

The current $j_z(g)$ originates from a 
conserved 4-dimensional current
corresponding to a chiral field transformation
\footnote{The important role of generalized currents corresponding to 
non-linear chiral field transformations 
was pointed out in 
\cite{chiralring} where generalized Konishi anomalies 
are used to determine the quantum chiral ring.}
\begin{equation}
\delta \phi^i = g^i(\phi).
\label{genTransform}
\end{equation}
For non-linear functions $g^i$ 
the transformation (\ref{genTransform}) does not generate any global symmetries.
Therefore, in contrast to the currents corresponding to the global 
symmetries in general there is no obvious prescription for the duality map
of generalized currents in the untwisted theory.

An ordinary current is quadratic in fields of matter
$$j_\mu = \sum_i q_i( \bar{\phi}_i D_\mu\phi^i-D_\mu\bar{\phi}_i \phi^i
+\bar{\psi}_i\sigma_\mu\psi^i),$$
where $q_i$ are the charges of the chiral fields $\phi^i$ and
corresponds to $j_z(g)$ with linear functions $g^i$.
In general, the central charges of the conserved flavor currents 
are not invariant under the renormalization group flow
\footnote{However, in many cases the values of the central functions $b$
can be computed at the infrared limit, e.g. at the conformal fixed points \cite{central}.}
$$\langle j_{\mu}(x)j_{\nu}(0)\rangle={1\over 16\pi^4}(\delta_{\mu\nu}\partial^2-
\partial_{\mu}\partial_{\nu})
\left({b(e(1/x))\over x^4}\right),$$ 
where $e$ stands for coupling constants.
This fact does not contradict the invariance of the OPE of $j_z(g)$.
The difference between $j_z(g)$ and the $z$-component of 
$j_\mu$ is proportional
to $\partial_z(\bar{\phi}_i g^i)$, and $\bar{\phi}_i g^i$ is the lowest component of
the supermultiplet containing the current $j_\mu$.
This difference is responsible for the different renormalization properties.

Following \cite{algebras} we also define a spin 2 current related to the stress tensor
\begin{equation}
T_{zz}=T_{zz}^{gauge}+T_{zz}^{matter}+\partial_zj_z(g),
\label{spintwo}
\end{equation}
where 
$$T_{zz}^{gauge}={2\pi \over e^2}\int_{\Sigma_2}{\rm Tr}[g^{w\bar{w}}
F_{zw}F_{z\bar{w}}-iD_z\bar{\lambda}\chi_z],$$
$$T_{zz}^{matter}=2\pi\int_{\Sigma_2}-D_z\bar{\phi}_iD_z\phi^i+
g^{z\bar{z}}\bar{\psi}_{zw,i}D_z\psi^i_{\bar{w}}.$$
By imposing the condition of $Q$-closeness $\{Q,T_{zz}\} = 0$
and using equations of motion one gets
\begin{equation}
W+\sum_jg^j\partial_jW=0.
\label{TW}
\end{equation}
The condition of cancellation of anomaly gives
\begin{equation}
-C_2(G)+\sum_i T(r_i)(1+2\partial_i g^i)= 0.
\label{virAnomaly}
\end{equation}
Various solutions to these constraints differ by a $z$-derivative of
a holomorphic current
corresponding to a conserved current of the untwisted theory.
A particular solution is 
\begin{equation}
g^i=-{1\over 2}R_i,
\label{KR}
\end{equation}
where $R_i$ stands for the $R$-charge of the superfield $\Phi_i$.
The operator $T_{zz}$ generates a Virasoro algebra in the $Q$-cohomology \cite{algebras}:
\begin{equation}
T(z,\bar{z})T(w,\bar{w})={c\over 2(w-z)^4}+{2T\over (z-w)^2}+
{\partial_z T\over z-w}+{\rm regular~terms}+\{Q, ...\}.
\label{TOPE}
\end{equation}
As discussed above the central charge $c$
of this Virasoro algebra can be computed by 
evaluating appropriate one-loop diagrams.
Moreover, the contributions of the non-zero modes 
are canceled due to the underlying supersymmetry.
The contributions of zero modes on $\Sigma_2$ in the denominators of the propagators
give:
\begin{equation}
c=-\chi{\rm dim}~G+2\sum_i {\rm dim}~r_i\left(1-3R_i+
{3 \over 2}R_i^2\right)~{\rm index}(R_i/2).
\label{vir}
\end{equation}
The first term in eq. (\ref{vir}) corresponds to the contribution of the gauge sector
\cite{algebras} while the second one comes from diagrams with propagators of
field of matter.
It is expedient to rewrite
the expression for the central charge as follows
\begin{equation}
c=\chi\left[-{\rm dim}~G +
\sum_i {\rm dim}~r_i \left(1-3R_i+{3\over 2}R_i^2\right)(1-R_i)\right]=
\label{virResult}
\end{equation}
$${\chi\over 2}[-2{\rm dim}~G+\sum_i(3 (1-R_i)^3 -
(1-R_i))]= {\chi\over 2}(-3\langle R^3\rangle + \langle R\rangle).$$
Here $\langle R^3\rangle$ and $\langle R\rangle$ stand for the corresponding coefficients
for 't Hooft anomalies for $U(1)^3_R$ and $U(1)_R$ respectively.
For a different choice of $j_z(g)$ in eq. (\ref{spintwo}) the value
of the central charge generically would involve other 't Hooft 
anomalies.
Notice that eq. (\ref{virResult}) is proportional to the infrared value of 
the $a$ central charge of the stress tensor provided 
the twisting $R$-current belongs to the $N=1$ superconformal multiplet at the
fixed point of the renormalization group \cite{anselmiEFJ} 
(for a recent discussion see \cite{kutasovLast}).

\section{Electric-magnetic duality}

In this section we apply the above construction to
theories related by electric-magnetic duality.
Since the coefficients in the OPE's of the holomorphic currents are invariant under 
renormalization group flows the algebras generated by the OPE's
have to be isomorphic in electric and magnetic theories.
In particular, the central charges have to coincide in the theories related 
by duality.
This isomorphism provides an additional test for electric-magnetic 
dualities.
However, in order to establish such an isomorphism one has
to specify a map between holomorphic currents 
in the theories related by duality.

\subsection{$SU(N_c)$ theory with fundamental flavors}

Consider the $SU(N_c)$ supersymmetric gauge theory with $N_f$
flavors of the chiral superfields 
$Q_i$ and $\tilde{Q}^i$, $i=1,...N_f$ in
the fundamental and antifundamental representation respectively, $N_f \geq N_c$. 
The anomaly free global symmetry is
\begin{equation}
G_f = SU(N_f)\times SU(N_f)\times U(1)_B \times U(1)_R.
\label{globalGroup}
\end{equation}
The quark fields transform under (\ref{globalGroup}) as
$$Q:~\left(N_f,1,1,1-{N_c\over N_f}\right),~~~
\tilde{Q}:~\left(\bar{N_f},-1, 1-{N_c\over N_f}\right).$$
The chiral ring is generated by the following gauge invariant operators
$$M^i_j = Q^i\tilde{Q}_j,$$
$$B^{[i_1...i_{N_c}]} = Q^{i_1}...Q^{i_{N_c}},$$
$$\tilde{B}_{[j_1...j_{N_c}]} = \tilde{Q}_{j_1}...\tilde{Q}_{j_{N_c}}.$$
In the infrared limit this theory has a dual (magnetic) description \cite{emseiberg}
in terms of the $SU(N_f-N_c)$ supersymmetric theory with 
$N_f$ fundamental flavors $q_i$ and $\tilde{q}^i$, $i = 1,...,N_f$, and 
a color-singlet meson field $M_i^j$, $i,j = 1,...,N_f$.
The magnetic theory has a superpotential 
$$W_{mag}=q_iM^i_j\tilde{q}^j.$$
The fields in the magnetic theory transform as
$$q:~\left(\bar{N_f},1,{N_c\over N_f-N_c},{N_c\over N_f}\right),$$
$$\tilde{q}:~\left(1,N_f,{N_c\over N_f-N_c},{N_c\over N_f}\right),$$
$$M:~\left(N_f,\bar{N_f},0,2(1-{N_c\over N_f})\right).$$
The chiral ring is generated by $M^i_j$ and baryons defined similarly with
the electric theory.

In the twisted electric theory 
one can define a holomorphic current corresponding to the baryon 
symmetry $U(1)_B$
$$j_z^{B}[Q,\tilde{Q}]=2\pi\int_{\Sigma_2}d^2w
\sum_{i=1}^{N_f}[({-\partial_z}\bar{Q}_iQ^{i}+
\bar{\psi}_{i,zw}\psi_{\bar{w}}^{i})-
({-\partial_z}\bar{\tilde{Q}}^{i}\tilde{Q}_i+
\bar{\tilde{\psi}}_{zw}^{i}\tilde{\psi}_{i,\bar{w}})]$$
where we use the same notations for the bosonic components of 
the chiral supermultiplets $Q^i$ and $\tilde{Q}_i$ as for 
the corresponding supermultiplets, and 
$\psi^{i}_{\bar{w}}$, $\bar{\psi}_{i,zw}$ 
($\tilde{\psi}_{i\bar{w}}$, $\bar{\tilde{\psi}}^i_{zw}$) 
stand for fermionic superpartners of $Q^i$($\tilde{Q}_i$) 
respectively. 
In a similar way, one can define the baryon current in the magnetic theory
$${N_c\over N_f-N_c}j_z^{B}[q, \bar{q}].$$
The coefficient $N_c/(N_f-N_c)$ is introduced for the correct
normalization of the current. 

It is easy to check that these currents are $Q$-closed and 
holomorphic in the $Q$-cohomology (see Eqs. (\ref{Qclose}) and (\ref{holom})).
Moreover, they act on the baryon operators in the same way in the both
theories and 
correctly compute the baryon charges
$$j_z^B(z,\bar{z})~B^{[i_1...i_{N_c}]}(u, \bar{u})=
-{1\over 2(z-u)^2}N_cB^{[i_1...i_{N_c}]}(u, \bar{u})+...$$
Therefore, we conclude
$$j_z^{B}[Q,\tilde{Q}]=j_z^{B}[q,\tilde{q}]+Q{\rm -exact~terms}.$$
As discussed in the previous section 
the central charge of this current is proportional to the 
corresponding 't Hooft anomaly $\langle U(1)_B^2U(1)_R\rangle=-2N_c^2$
that has the same value in both theories.
In a similar way, one can identify the $SU(N_f)$ and 
the spin-2 $T_{zz}$ currents in the 
electric and magnetic theories and check the matching of the corresponding
central charges.

\subsection{$SU(N_c)$ theory with fundamental flavors and a chiral superfield in adjoint representation}

The Kutasov-Schwimmer model
\cite{kutasov}, \cite{kutsch} is more interesting for study of holomorphic currents.
In this case, one has a $SU(N_c)$ gauge theory with 
$N_f$ fundamental flavors, $Q^i$, $\tilde{Q}_i$, $i=1,...,N_f$ and
a chiral superfield $X$ in the adjoint representation of 
the gauge group with a superpotential 
\begin{equation}
W_{el}={s_0\over k+1}{\rm Tr}X^{k+1},~~k\geq 1,
\label{elSP}
\end{equation}
where $s_0$ is a complex parameter, $kN_f \geq N_c$.
The anomaly free global symmetry group given by (\ref{globalGroup}).
The matter fields transform under (\ref{globalGroup}) as:
$$Q:~~\left(N_f,1,1,1-{2\over k+1}{N_c\over N_f}\right),$$
$$\tilde{Q}:~~\left(1,\bar{N_f},1,1-{2\over k+1}{N_c\over N_f}\right),$$
$$X:~~\left(1,1,0,{2\over k+1}\right).$$
The dual (magnetic) theory has gauge group $SU(kN_f-N_c)$,
$N_f$ fundamental flavors, $q_i$, $\tilde{q}^i$, 
an adjoint field $Y$, and gauge singlets $M_j$, $j=1,...k$
with the superpotential
\begin{equation}
W_{mag}=-{s_0\over k+1}{\rm Tr}Y^{k+1}+
{s_0\over \mu^2}\sum_{j=1}^k M_j\tilde{q}Y^{k-j}q.
\label{magSP}
\end{equation}
The transformation properties of the magnetic matter fields 
under (\ref{globalGroup}) are:
$$q:~~\left(\bar{N_f},1,{N_c\over\bar{N_c}}, 
1-{2\over k+1}{\bar{N_c}\over N_f}\right),$$
$$q:~~\left(1,N_f,{N_c\over\bar{N_c}}, 
1-{2\over k+1}{\bar{N_c}\over N_f}\right)$$
$$Y:~~\left(1,1,0,{2\over k+1}\right),$$
$$M_j:~~\left(N_f,\bar{N_f},0,2-{4\over k+1}{N_c\over N_f}
+{2\over k+1}(j-1)\right)$$
where we denoted $\bar{N_c}=kN_f-N_c$.

The chiral rings and their duality map were studied in \cite{emmap}.
Following \cite{emmap} we define the baryon-like operators 
as follows
\begin{equation}
B^{i^{(1)}_1...i^{(1)}_{n_1},...,
i^{(k)}_1...i^{(k)}_{n_k}}=
\label{baryons}
\end{equation}
$$\epsilon^{\alpha^{(1)}_1...\alpha^{(1)}_{n_1},...,
\alpha^{(k)}_1...\alpha^{(k)}_{n_k}}
\left(Q^{i^{(1)}_1}_{\alpha^{(1)}_1}...Q^{i^{(1)}_{n_1}}_{\alpha^{(1)}_{n_1}}\right)...
\left((X^{k-1}Q)^{i^{(k)}_1}_{\alpha^{(k)}_1}...
(X^{k-1}Q)^{i^{(k)}_{n_k}}_{\alpha^{(k)}_{n_k}}\right),$$
where $\sum_i n_i=N_c$.
The duality map relating operators $B$ with the analogously 
defined dual baryon operators $\bar{B}$ is
\begin{equation}
B^{i^{(1)}_1...i^{(1)}_{n_1},...,i^{(k)}_1...i^{(k)}_{n_k}}=
P\left(\prod_{i=1}^k {1\over \bar{n}_i!}\right)
(-s_0)^{kN_f\over 2}(\mu^2)^{-\bar{N}_c\over 2}
\Lambda^{{k\over 2}(2N_c-N_f)}
\label{baryonMap}
\end{equation}
$$ \epsilon^{i^{(1)}_1...i^{(1)}_{n_1},t^{(k)}_1...t^{(k)}_{\bar{n}_k}}
...\epsilon^{i^{(k)}_1...i^{(k)}_{i_k},t^{(1)}_1...t^{(1)}_{\bar{n}_1}}
\bar{B}_{ t^{(k)}_1...t^{(k)}_{\bar{n}_1},...,
t^{(1)}_1...t^{(1)}_{\bar{n}_k}}$$
where $\bar{n}_l=N_f-n_{k+1-l}$, $l=1,...k$, 
$P^2=(-1)^{{k(k-1)\over 2}N_f-N_c}$, $\mu$ is an auxiliary scale parameter.
In addition to baryons the chiral ring contains mesons
$(M_j)^i_l=\tilde{Q}_lX^{j-1}Q^i$ and the generators
Tr$X^j$ (Tr$Y^j$), $j=2,...,k$.
Notice that from equations of motion we also have
\begin{equation}
X^k-{1\over N_c}{\rm Tr}X^k={\rm Q-exact}, 
~~~Y^k-{1\over N_c}{\rm Tr}Y^k={\rm Q-exact}.
\label{chiralRingConditions}
\end{equation}

Similarly with the discussion of $SU(N_c)$ SQCD
one can easily identify the holomorphic currents corresponding
to the baryon symmetry $U(1)_B$ and the spin-2 currents $T_{zz}$.
The associated central charges match in the electric and magnetic theories
since they are proportional to appropriate 't Hooft anomalies.

A new element that appears in the Kutasov-Schwimmer model 
is an additional 
structure that imposes non-trivial conditions on the 
duality map of the chiral ring. 
In the case when $\Sigma_2$ is a torus $T^2$ with a flat metric one can also define
{\it generalized} baryon currents \footnote{The integrand in this equation is not a $(1,1)$-form 
on an arbitrary $\Sigma_2$. On a torus this is not important since it can be adjusted by using 
a constant 1-form.}
\begin{equation}
j_z^j[Q,\tilde{Q},X]=2\pi\int_{T^2}
(-D_z\bar{Q}_i X^jQ^i+
\bar{\psi}_i X^j\psi_{\bar{w}}+
\bar{\psi}_ijX^{j-1}\chi_{\bar{w}}Q^i)-
(Q^i\to \tilde{Q}_i,~\psi^i\to\tilde{\psi}_i).
\label{genBaryonCurrent}
\end{equation}
Here $\chi_{\bar{w}}$ stands for the fermionic superpartner of $X$.
Similarly we define generalized baryon current 
$j_z^j[q,\tilde{q},Y]$ on the magnetic side.
Since the Euler characteristic 
vanishes for torus the central terms in the OPE of these currents also vanish.

As discussed in the previous section, these currents act as vector fields
on baryon operators $B$ and $\bar{B}$.
In what follows we study consistency of this action with eq. (\ref{baryonMap})
and the duality map determined in \cite{emmap}.

Let us denote $B$ defined in eq. (\ref{baryons}) as
$B[n_1,...n_k]$.
Consider the action of $j_z^j[Q,\tilde{Q},X]$ on an arbitrary factor
in eq. (\ref{baryons})
\begin{equation}
(X^{l}Q)^{s}_{\lambda}\to (X^{l+j}Q)^{s}_{\lambda}.
\label{element}
\end{equation}
If $l+j < k$ this action corresponds to
\begin{equation}
B[n_1,..,n_l,..,n_{l+j},...,n_k]\to 
B[n_1,..,n_l-1,..,n_{l+j}+1,...,n_k].
\label{baryonElement}
\end{equation}
Thus, the baryon 
$B[n_1,...,n_k]$ transforms into 
a sum of baryons with a different set of $n_l$ under the action of $j_z^j[Q,\tilde{Q},X]$.
For consistency with eq. (\ref{baryonMap}),
according to eq. (\ref{baryons}) one should have in the magnetic theory
$$\bar{B}[\bar{n}_1,..,
\bar{n}_{k+1-l-j},..,\bar{n}_{k+1-l},...,\bar{n}_k]\to 
\bar{B}[\bar{n}_1,..,\bar{n}_{k+1-l-j}-1,..,
\bar{n}_{k+1-l}+1,...,\bar{n}_k].$$
Indeed, one can see that 
$j_z^j[q,\tilde{q},Y]$ acts on a factor
$(X^{k+1-l-j}Q)^{s}_{\lambda}$ as 
\begin{equation}
(Y^{k+1-l-j}q)^{s}_{\lambda}\to (Y^{k+1-l}q)^{s}_{\lambda}.
\label{elementY}
\end{equation}
We will show that this action produces $\bar{B}$ operators that exactly match 
eq. (\ref{baryonElement}).

First, one has to check that the numbers of baryon operators 
and phase factors match
for $B$ and $\bar{B}$ on the right hand side.
Let us denote by $[n_l]$ ($[\bar{n}_l]$) 
the set of flavor indices that are assigned to
factors $X^lQ$ ($Y^lq$) in the electric (magnetic) theory.
Notice that the transformation (\ref{element}) for a factor with a flavor index in $[n_l]\cap [n_{l+j}]$ 
does not produce any baryon operator because 
of antisymmetrization of the color indices in eq. (\ref{baryons}).
We denote the set of available flavor indices in $[n_l]$ (i.e. indices corresponding 
to the factors that are not annihilated by (\ref{element})) as $[\nu_l]=[n_l]-([n_l]\cap [n_{l+j}])$, 
and $\nu_l$ is the number of
elements in $[\nu_l]$. 
Obviously, $\nu_l\leq n_l$.
On the magnetic side the set of flavor indices in $[\bar{n}_{k+1-l-j}]$ 
may also overlap with that of $[\bar{n}_{k+1-l}]$.
The action of $j_z^j[q,\tilde{q},Y]$ on the factors 
with flavor indices in $[\bar{n}_{k+1-l-j}]\cap [\bar{n}_{k+1-l}]$
annihilates the operator $\bar{B}$.
The actual number of produced baryon operators equals $\nu_l$.
An easy way to see this is to consider the reduced sets
$[n'_l]=[n_l]-[n_{l+j}]$, $[n'_{l+j}]=[n_{l+j}]-[n_l]$,
$[\bar{n'}_{k+1-l}]=[\bar{n}_{k+1-l}]-[\bar{n}_{k+1-l-j}]$
and $[\bar{n'}_{k+1-l-j}]=[\bar{n}_{k+1-l-j}]-[\bar{n}_{k+1-l}]$.
Since $[n'_l] \cap [n'_{l+j}]=
[\bar{n'}_{k+1-l}]\cap [\bar{n'}_{k+1-l-j}]=\emptyset$
and 
$[n'_l]\cup [n'_{l+j}]=[\bar{n'}_{k+1-l}]\cup [\bar{n'}_{k+1-l-j}]$
we have $[n'_l]=[\bar{n'}_{k+1-l-j}]$.
Hence, the set of available flavor indices on the electric side coincides with those
on the magnetic side.
Thus, under the action of $j_z$ we get equal numbers of baryons 
in the electric and magnetic theories.

Let us analyze how the flavor indices map for each of the baryons.
Originally on the magnetic side we had $\epsilon$ factors
carrying flavor indices
\begin{equation}
...
~\epsilon^{i_1^{(l+j)}...i_{n_{l+j}}^{(l+j)},
t_1^{(k+1-l-j)}...p...t_{\bar{n}_{k+1-l-j}}^{(k+1-l-j)}}
~...
~\epsilon^{i_1^{(l)}...p...i_{n_{l}}^{(l)},
t_1^{(k+1-l)}...t_{\bar{n}_{k+1-l}}^{(k+1-l)}}
~...
\label{epsilonFactors}
\end{equation}
Let us consider the action of $j_z[q,\tilde{q},Y]$ on the factor 
$Y^{k+1-l-j}q$ with a flavor index $p$.
Under this action the index $p$ 
moves from the set of magnetic flavor indices
$[\bar{n}_{k+1-l-j}]$ into the set of electric flavor indices 
$[n_{l+j}]$, and simultaneously it moves from 
$[n_{l}]$ into $[\bar{n}_{k+1-l}]$.
This exactly corresponds to what happens with the factor
$X^lQ$ with the flavor index $p$.
Since a baryon operator is uniquely defined by specifying the corresponding 
sets of flavor indices, we see that the flavor index structure in the 
relation (\ref{baryonMap}) is supported by this action.

Now let us compute the sign factors that are due to 
a rearrangement of color and flavor indices after the action of
$j_z[q,\tilde{q},Y]$.
In order to simplify the discussion, let us choose the convention
that the unsummed flavor indices in eq. (\ref{baryons})
are in the ascending order.
It is easy to see that due to the rearrangement of color 
indices on the electric side 
the produced baryon acquires a factor $(-1)^{P_{el}}$,
where
$$P_{el}=n_l-i^{(l)}(p)+i^{(l+j)}(p)-1+\sum_{r=l+1}^{l+j-1}n_r.$$
Here $i^{(l)}(p)$ stands for the position of the index $p$ in the
ordered set of indices $i^{(l)}$.
For the corresponding baryon on the magnetic side one has a factor
$(-1)^{P_{mag}}$ where
$$P_{mag}=\left(\bar{n}_{k+1-l-j}-t^{(k+1-l-j)}(p)+
t^{(k+1-l)}(p)-1+\sum_{r=k+1-l-j+1}^{k+1-l-1}\bar{n}_r\right)+$$
$$\left(n_{l+j}-i^{(l+j)}(p)+t^{(k+1-l-j)}(p)-1+
n_l-i^{(l)}(p)+t^{(k+1-l)}(p)-1\right).$$
The terms in the first bracket come from the rearrangement of 
the color indices while the second bracket is due to the rearrangement of
flavor indices in the $\epsilon$ factors (\ref{epsilonFactors}).
Thus, the relative factor 
\begin{equation}
(-1)^{P_{el}-P_{mag}}=(-1)^{jN_f}
\label{relativeFactor}
\end{equation}
depends only on $j$ but not on the details of the index rearrangements.
This factor can be reproduced if we include 
\begin{equation}
(-1)^{N_f\sum_l ln_l}
\label{inP}
\end{equation}
into the factor $P$ in
the baryon operator mapping (\ref{baryons}).
We will show that this assumption agrees with the mass flow
analysis in \cite{emmap} (section (5.1)).

As pointed out in \cite{emmap} there is a sign ambiguity the duality map in eq. (\ref{baryonMap})
due to non-anomalous global symmetries (therefore the authors of \cite{emmap} focused on
phase factor for $B\tilde{B}$ that is invariant).
However, relative phases of baryons with the same flavor indices does not change under
phase rotations of quarks superfields.
The factor (\ref{inP}) controls such relative phases. 
By introducing a mass term $m\tilde{Q}Q$ in the electric theory the number 
of flavors reduces $N_f\to N_f-1$,
and only baryons without a flavor index $N_f$ survive in the infrared limit.
This flow corresponds \cite{emmap} to Higgsing $SU(\bar{N}_c)\to SU(\bar{N}_c-k)$ 
in the magnetic theory with non-zero vevs
\begin{equation}
\tilde{q}^{N_f}_\alpha=\delta_{\alpha, 1}\left({m\mu^2\over -s_0}\right)^{1\over k+1},~~
q^\alpha_{N_f}=\delta^{\alpha,k}\left({m\mu^2\over -s_0}\right)^{1\over k+1} ,~~
Y^\alpha_\beta=\delta^\alpha_{\beta-1}\left({m\mu^2\over -s_0}\right)^{1\over k+1},
\label{higgsing}
\end{equation}
and hence, the number of flavors in the magnetic theory also reduces $N_f\to N_f-1$.
In the expression for a magnetic baryon $\bar{B}$ the broken color indices 
$\alpha=1,...k$ are taken by 
the factors $Y^jq^{N_f}$.
The rearrangements of color and flavor indices that is needed 
to bring baryons to the standard form, give the factors
$$(-1)^{{k(k-1)\over 2}+(N_f-1){k(k+1)\over 2}+\sum_l ln_l}
~~{\rm and}~~(-1)^{(N_f-1){k(k+1)\over 2}+\sum_l ln_l}$$
for $\bar{\tilde{B}}[\bar{n}_1,...\bar{n}_k]$ and $\bar{B}[\bar{n}_1,...\bar{n}_k]$ respectively.
Thus, the change of the phase in the map for $B\tilde{B}$ is 
$(-1)^{k(k-1)\over 2}$ as stated in \cite{emmap}.
The factor $(-1)^{\sum_l ln_l}$ is exactly equal to the change of (\ref{inP}) under $N_f\to N_f-1$.

If $l+j \geq k$ eq. (\ref{element}) is equivalent to 
\begin{equation}
(X^{l}Q)^{s}_{\lambda}\to (X^{l+j-k}Q)^{s}_{\lambda}~
{1\over N_c}{\rm Tr}X^k
\label{elementHigh}
\end{equation}
and hence,
$$B[n_1,..,n_{l+j-k},..,n_l,...,n_k]\to 
B[n_1,..,n_{l+j-k}+1,..,n_l-1,...,n_k]~{1\over N_c}{\rm Tr}X^k.$$
Here the numbers $n_l$ jump under the action of $j_z^j[Q,\tilde{Q},X]$.
Similarly with the previous case, the numbers of produced baryons and phase factors
match in the electric and magnetic theories.
The only new element is the presence of different factors 
${1\over N_c}{\rm Tr}X^k$ and ${1\over \bar{N}_c}{\rm Tr}Y^k$ 
in the formulas for $B$ and $\bar{B}$ transformations respectively.
However, according to the results of \cite{emmap}
in the chiral ring one has
\begin{equation}
{1\over N_c}{\rm Tr}X^k={1\over \bar{N}_c}{\rm Tr}Y^k.
\label{kmap}
\end{equation} 
Thus, the currents $j_z^j[Q,\tilde{Q},X]$ and $j_z^j[q,\tilde{q},Y]$ 
act in the same way on the baryon operators up to a sign factor.
These currents annihilate
the meson operators and Tr$X^l$, Tr$Y^l$.
Thus, we conclude that
\begin{equation}
j_z^j[Q,\tilde{Q},X]=j_z^j[q,\tilde{q},Y]+
Q-{\rm exact~terms}.
\label{currMap}
\end{equation}
Notice that the identity (\ref{kmap})
obtained in \cite{emmap} from very different arguments 
was crucial for the consistency of
this identification.
One can view this fact as an additional argument supporting
the duality picture developed in \cite{emmap}.

One can extend this analysis to the currents 
\begin{equation}
j_z^{j,a}[Q,\tilde{Q},X]=2\pi\int_{T^2}
(-D_z\bar{Q}T^a X^jQ+
\bar{\psi}T^a X^j\psi_{\bar{w}}+
\bar{\psi}T^ajX^{j-1}\chi_{\bar{w}}Q).
\label{genBaryonCurrentNA}
\end{equation}
those are not singlets
with respect to the $SU(N_f)\times SU(N_f)$ flavor group.
Here $T^a$ stand for a generator of $SU(N_f)$, 
and flavor and color indices are implicit.
This current corresponds to the following transformations 
\begin{equation}
\delta_j^a Q\to T^aX^jQ.
\label{sunfel}
\end{equation}
Let us determine the corresponding transformation on the magnetic side.
The action on the magnetic quarks reads as
\begin{equation}
\delta_j^a q\to T^aY^jq
\label{sunfmagQuarks}
\end{equation}
while for mesons one has
\begin{equation}
\delta_j^a M_l=-T^a M_{l+j},~~l+j\leq k,
\label{sunfmagMesons}
\end{equation}
$$\delta^a_j (M_l)^{f'}_f = - (M_{l+j-k})^{f'}_p(N_{2k-l}N_{k-l}^{-1})^p_f,~~l+j>k,$$
where in the second line we show explicitly the $SU(N_f)\times SU(N_f)$ flavor $f, f', p$ indices and 
$$(N_l)^{f'}_f\equiv\tilde{q}^{f'}Y^lq_f.$$
Notice that in the chiral ring the transformation at $l+j>k$ reduces to
$$\delta_j^a M_l\to-{1\over \bar{N}_c}{\rm Tr}Y^k T^a M_{l+j-k},~~l+j\geq k+1.$$

Thus, the total current  reads as
\begin{equation}
j_{tot,z}^{j,a}[q, \bar{q},M,\bar{M},Y]=j_z^{j,a}[q,\tilde{q},Y]+m_z^{j,a}[M,\bar{M},Y],
\label{genMagBaryonCurrent}
\end{equation}
where 
\begin{equation}
m_z^{j,a}[M,\bar{M},Y]=-2\pi\int_{T^2}(-\sum_{l} \partial_z\bar{M_l}\delta^a_j M_l+
{\rm fermionic~terms}).
\label{genMesonCurrent}
\end{equation}
The fermionic terms in eq. (\ref{genMesonCurrent}) can be restored by using eq. (\ref{J}).
Clearly, the action of $j_{tot,z}^{j,a}[q, \bar{q},M,\bar{M},Y]$ 
on meson operators supports the duality map $M_l=\tilde{Q}X^lQ$.

The problem with eq. (\ref{genMagBaryonCurrent}) is that 
the terms corresponding to $l+j >k$ are singular in chiral fields.
This fact apparently implies that one has to extend the ultraviolet off-shell formulation of magnetic theory
by introducing additional chiral superfields.
These additional degrees of freedom should become massive in the infrared limit.
We will assume that the current (\ref{genMagBaryonCurrent})
is well defined in the context of operator product expansion.

Consider the action of $j_z^{j,a}[Q,\tilde{Q},X]$ on the baryon
$B[n_1,...,n_k]$.
For a diagonal generator $T^a$ the analysis is similar to the case of the singlet current.
Therefore, it is sufficient to
consider only non-diagonal generators $(S^{(p,q)})_{ij} = \delta_{ip}\delta_{jq}+
\delta_{jp}\delta_{iq}$ and 
$(A^{(p,q)})_{ij}=i\delta_{ip}\delta_{jq}-i\delta_{jp}\delta_{iq}$, $p\neq q$.
Consider a factor $X^lQ^p$ in $B[n_1,...,n_k]$.
The action of a non-diagonal current on it may be decomposed into the action of 
a diagonal current and a change of the index $p\to q$.
Assuming that this action does not annihilate the baryon operator one has
$p\in [n_l]$ and $q\notin [n_{l+j}]$.
Hence, $p\notin [\bar{n}_{k+1-l}]$ and $q\in [\bar{n}_{k+1-l-j}]$.
Thus, moving the index $p$ from $[n_l]$ to $[n_{l+j}]$
and replacing it with $q$  
corresponds to moving the index $q$ from $[\bar{n}_{k+1-l-j}]$ to $[\bar{n}_{k+1-l}]$
and replacing it with $p$.
This is obviously a one-to-one correspondence that supports the relation (\ref{baryonMap}).
Therefore, we have the same sets of baryons produced in the electric and magnetic theories.

The analysis of the phase factors is similar to the case of diagonal currents.
The only subtlety here is that the currents corresponding to the generators
$A^{(p,q)}$ produce different
additional factors $i$ and $-i$ on the electric and magnetic side respectively.
This difference in sign is accounted for by the fact that the electric and magnetic
quarks belong to the $N_f$ and $\bar{N}_f$ representations of the group $SU(N_f)$
respectively.
Thus, we have
\begin{equation}
j_z^{j,a}[Q,\tilde{Q},X]=j_{tot,z}^{j,a}[q, \bar{q},M,\bar{M},Y]+
Q-{\rm exact~terms}.
\label{currSUMap}
\end{equation}
Hence, the algebras generated by the holomorphic currents match 
in the electric and magnetic theories.

\section{Discussion}

In this paper we studied the algebra generated by non-local holomorphic currents
in twisted supersymmetric theories.
The central charge in the OPE of 
the holomorphic spin-2 current with itself
coincides with the infrared value of the $a$ central charge, i.e. the coefficient of the
Euler density in the trace anomaly.
Holomorphic spin-1 currents related to the global symmetries have central charges
proportional to 't Hooft anomalies.
We also discussed the algebra of generalized holomorphic currents corresponding to
non-linear transformations of chiral fields.

The holomorphic algebras have interesting applications in the context of electric-magnetic duality.
In particular, central charges of holomorphic currents in the electric theory should 
coincide with the corresponding central charges in the magnetic theory.
Matching generalized holomorphic currents in the electric and magnetic theories 
imposes non-trivial
conditions on the duality map of the chiral ring. 
We considered in some detail the $SU(N_c)$ supersymmetric QCD and 
the adjoint supersymmetric QCD (Kutasov-Schwimmer models),
and found agreement of the holomorphic structure with the electric-magnetic duality.
However, we found that in the magnetic description the generalized non-abelian flavor currents are not polynomial.

This analysis can be generalized to other models that exhibit electric-magnetic duality.
An interesting example is the adjoint supersymmetric QCD discussed above
with vanishing superpotential.
As discussed in \cite{emmap}, when $N_f\lesssim 2N_c$ the theory flows to 
a non-Abelian Coulomb phase in the infrared.
There is a one parameter set of $R$-symmetries. 
The physical $R$-symmetry can be determined by using the results of
\cite{intriligatorWecht}.
The renormalization group flows and phase structure of this theory has been studied in details
in \cite{kutasovLast}.
However, there is no known magnetic description.

This theory can be twisted by using the $R$-current that corresponds to the following 
$R$-charges of the chiral fields:
$R(X)=0,~R(Q)=0$ and $R(\tilde{Q})=1$.
Since there is no superpotential the chiral ring is generated by 
Tr$X^j$, $j=0,...$ and baryon operators defined similarly with eq. (\ref{baryons}).
However, in this case there is no restrictions on powers of $X$.
For this choice of the twisting $R$-current the field $X$ has zero spin, and
the generalized holomorphic currents 
$j_z^j[Q,\tilde{Q},X]$ can be defined for arbitrary surface $\Sigma_2$.
These holomorphic currents generate an {\it infinite} algebra with non-trivial "central" 
operators Tr$X^j$.
If a magnetic description of the theory exists, the magnetic theory has to possess
a holomorphic algebra isomorphic to that on the electric side. 
Then this isomorphism may give information on a duality map for the chiral ring.

Notice that in models where {\it accidental} symmetries appear in the infrared limit
the actual holomorphic algebra is larger than that visible in the ultraviolet regime.
This happens when some chiral operators reach the unitarity
bound under the renormalization group flow and decouple.
For example, an accidental symmetry corresponding to decoupling of $Q\tilde{Q}$
appears in the presence of a superpotential 
(\ref{elSP}) for $k=2$ and $N_f<N_c$ in the Kutasov-Schwimmer model.
Models with accidental symmetries may be an interesting playground for 
applications of holomorphic currents.

It would be also interesting to generalize the construction of holomorphic 
algebras to deformed models where $R$-symmetry is broken by 
a superpotential.
By promoting the coupling in the superpotential to chiral superfields
one can still twist the theory.
Such a generalization could allow considering the behavior
of holomorphic currents under interpolation between various conformal theories.
Also, by integrating over massive fields one can get 
an effective sigma-model for the coupling constants
where the holomorphic algebra should also have a realization.
By matching the holomorphic algebras in these two limits
could obtain some conditions on the quantum metric on the space of deformations
of the model.

\acknowledgments{I would like to thank M. Bershadsky for useful discussions.}

\end{document}